\documentclass[final,5p,times,twocolumn]{elsarticle}
\biboptions{sort&compress}
\usepackage{amsmath}
\usepackage{makeidx}
\usepackage{amsfonts}
\usepackage[ansinew]{inputenc}
\usepackage[usenames,dvipsnames]{pstricks}
\usepackage{subfigure}
\usepackage{epsfig}
\usepackage{pst-grad} % For gradients
\usepackage{pst-plot} % For axes
\usepackage[colorlinks,hyperindex]{hyperref}
\usepackage{booktabs}
\usepackage{multirow}
\usepackage{chngpage}
\usepackage{tabularx}
\usepackage{doi}
\usepackage{caption} 
\usepackage{xcolor}
\hypersetup
{
   colorlinks,%
   citecolor=blue,%
   linkcolor=blue,%
   urlcolor=blue,%
}

\mathchardef\mhyphen="2D
\makeindex

\begin{document}

\title{The evidence of $N=16$ shell closure and $\beta$-delayed neutron emission from $^{25}$F}

\author[utk]{J.~F.~Peltier}
\ead{peltierjack1@gmail.com}
\author[utk]{Z.~Y.~Xu}%
\author[utk]{I.~Cox}%
\author[utk]{R.~Grzywacz}%
\author[frib]{R.~S.~Lubna}%
\author[cns]{N.~Kitamura}%
\author[llnl]{S.~Neupane}%

\author[ornl]{J.~M.~Allmond}%
\author[utk]{J.~Christie}%
\author[frib,msu]{A.~A.~Doetsch}%
\author[utk]{P.~Dyszel}%
\author[misu,frib]{T.~Gaballah}%
\author[ornl]{T.~T.~King}%
\author[llnl]{K.~Kolos}%
\author[frib,msu]{S.~N.~Liddick}%
\author[utk]{M.~Madurga}%
\author[misu]{T.~H.~Ogunbeku\fnref{fn0}}%
\fntext[fn0]{Current affiliation: Lawrence Livermore National Laboratory, Livermore, California 94550, USA}
\author[frib,msu]{B.~M.~Sherrill}%
\author[utk]{K.~Siegl}%

\address[utk]{Department of Physics and Astronomy, University of Tennessee,
Knoxville, Tennessee 37996, USA}
\address[frib]{Facility for Rare Isotope Beams, Michigan State University,
East Lansing, Michigan 48824, USA}
\address[cns]{Center for Nuclear Study, University of Tokyo, Wako, Saitama
351-0198, Japan}
\address[llnl]{Lawrence Livermore National Laboratory, Livermore, California
94550, USA}
\address[ornl]{Physics Division, Oak Ridge National Laboratory, Oak Ridge,
Tennessee 37831, USA}
\address[msu]{Department of Physics and Astronomy, Michigan State University,
East Lansing, Michigan 48824, USA}
\address[misu]{Department of Physics and Astronomy, Mississippi State University,
Mississippi State, Mississippi 39762, USA}

\date{\today}

\begin{abstract}
   We measured the $\beta$-delayed neutron emission from $^{25}$F for the first time at the Facility for Rare Isotope Beams (FRIB). Using combined neutron and $\gamma$-ray detector systems of the FRIB Decay Station Initiator (FDSi), we observed $\beta$-decay transitions populating neutron unbound states between 4.2 and 8 MeV in $^{25}$Ne. The experimental results led to the revision of the $\beta$-decay half-life and $\beta$-delayed neutron-emission probability of $^{25}$F. The $\beta$-decay strength distribution of $^{25}$F extracted from the data agrees with the shell-model predictions using the USDB and SDPF-M effective interactions. This result indicates that the spherical neutron $N = 16$ shell gap persists in $^{25}$F and $^{25}$Ne.
\end{abstract}

% \begin{keyword}
   % keyword1 \sep keyword2 \sep keyword3
% \end{keyword}

\maketitle

\noindent\textbf{\textit{Introduction}---}Tracking and understanding shell evolution as a function of neutron or proton numbers is one of the most crucial tasks in modern nuclear physics. Nuclei with closed proton and neutron shells, which are referred to as doubly magic nuclei, are particularly simple in the configurations of their ground state and lowest-lying excited states. Therefore,  their properties are the cornerstones of the development of nuclear models.

Oxygen-24 is one of the doubly magic nuclei far from the stability line due to the large proton and neutron shell gaps at $Z=8$ and $N=16$, respectively \cite{dbm24O,dbm24O_2,dbm24O_3}. While $Z=8$ is a traditional shell closure observed in stable and near stable nuclei, the neutron shell gap at $N=16$ emerges in neutron-rich nuclei \cite{neutron16,PhysRev.75.1969}. It is induced by the energy spacing between the neutron $0d_{3/2}$ and $1s_{1/2}$ orbitals, which is especially large when the proton $0d_{5/2}$ orbital is unoccupied, as it is in the oxygen isotopes \cite{dbm24O,neutron16,n-20_monte-carlo}. The shifting of neutron magic number from $N = 20$ to $N = 16$ near the neutron dripline in oxygen has attracted tremendous interest from both experimental and theoretical studies \cite{n-rich_shellgap_spin-isospin,SM_evolution_exotic_nuclei,neutron16, unstable_magic_number,Ne_N-16_or_N-14, SM_orbital_location,25O_n-16,25Ne}. Due to the attractive nucleon-nucleon interaction between the proton $0d_{5/2}$ and neutron $0d_{3/2}$ orbitals, the $N = 16$ shell gap is expected to reduce when protons are added to the $0d_{5/2}$ orbital. Indeed, such experimental evidence has been observed in $^{26}$Ne \cite{Ne_N-16_or_N-14} and $^{25}$F \cite{25Fpknockout,25Fvs24O}. In particular, the values of spectroscopic factors extracted from a proton-knockout reaction $^{25}$F$(p,2p)^{24}$O suggested a substantial reduction of about 4.2 MeV in $^{25}$F compared with $^{24}$O \cite{25Fvs24O}. This, however, contradicts the conclusions from other authors: for example, in Refs. \cite{PhysRevC.110.034323,25Ne}, the $N = 16$ shell gap is found to be preserved in $^{25}$F and $^{24}$F respectively. Since the amount of reduction is sensitive to the proton-neutron interaction, it is of great interest to reinvestigate the evolution of the $N = 16$ shell gap in the vicinity of $^{24}$O to better benchmark nuclear models in the region.

In an allowed Gamow-Teller (GT) transition, a neutron is transformed into a proton of the spin-orbital partner. Therefore, it is extremely sensitive to the proton and neutron occupation in a given shell and an efficient tool to probe the shell structure. In the present work, the $\beta$ decay of $^{25}$F was revisited using combined neutron and $\gamma$-ray spectroscopy. The energies of $\beta$-delayed neutron emissions from $^{25}$F were measured for the first time, allowing for the reconstruction of excitation energies and $\beta$-decay feeding probabilities to neutron unbound states in $^{25}$Ne. Compared to the previous measurement on the $^{25}$F decay \cite{25Ne}, the $\beta$-decay strength ($S_{\beta}$) distribution was substantially extended to 8 MeV in $^{25}$Ne, 4 MeV above the neutron separation energy ($S_n$, 4.18 MeV in $^{25}$Ne \cite{masstable}). The newly extended $S_{\beta}$ distribution was compared to shell-model (SM) calculations using USDB \cite{usdbshell} and SDPF-M \cite{sdpfm} effective interactions with good agreement, suggesting a moderate reduction of the $N = 16$ shell gap in the ground state of $^{25}$F. In addition, the half life ($t_{1/2}$) and $\beta$-delayed neutron-emission probability ($P_{n}$) of $^{25}$F were revised from experimental data, deviating from previously measured values \cite{25Fdecay,25FdecayGANIL,25Ne}.

\begin{figure}[ht]
   \includegraphics[width=0.5\textwidth]{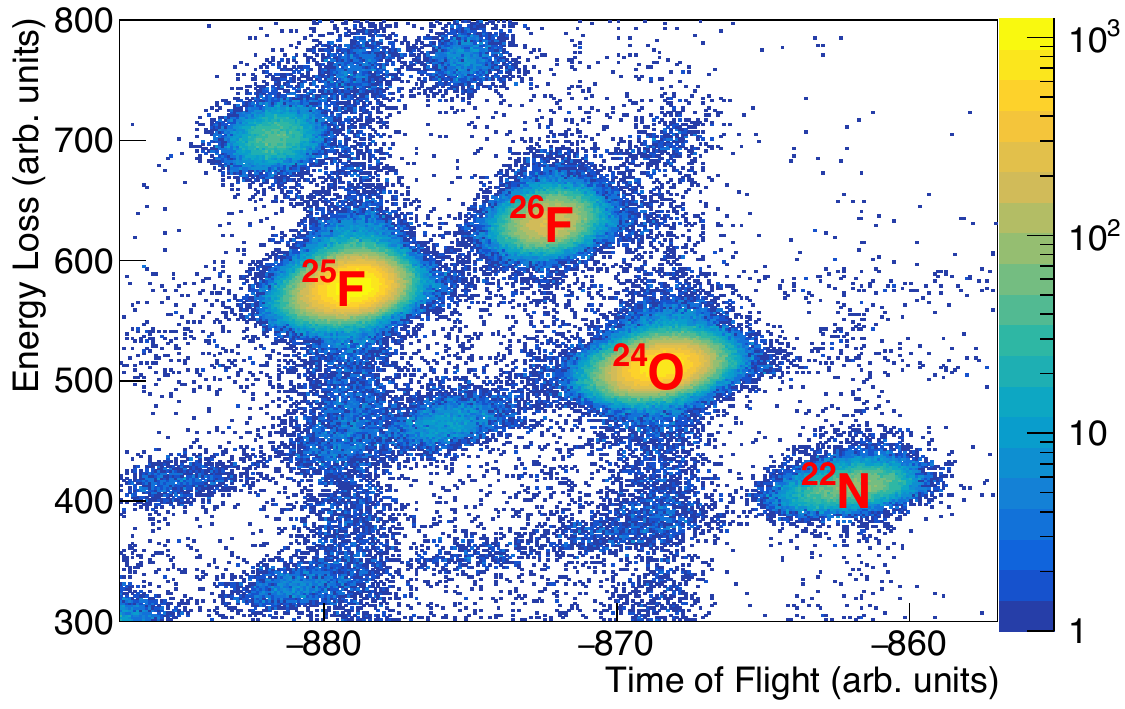}
   \caption{The particle identification plot from the beam time is shown, in which $^{22}$N, $^{24}$O, $^{25}$F and $^{26}$F are identified and labeled.}
   \label{fig:fig0}
   \end{figure}

\noindent\textbf{\textit{Experiment}---}The experimental data was gathered over the course of 25 hours of beam time at the Facility for Rare Isotope Beams (FRIB) using the FRIB Decay Station Initiator (FDSi). A 200 MeV/u $^{40}$Ar primary beam was impinged on an 8 mm thick carbon target with 5 kW total beam power. After $^{40}$Ar fragmentation, the resulting secondary beam was sent through the Advanced Rare Isotope Separator (ARIS) \cite{aris}, from which isotopes of interest were selected and identified on an event-by-event basis, and was delivered to the FDSi as a cocktail secondary beam. The time-of-flight (TOF) for each ion was recorded between a timing scintillator in ARIS and another in the diagnostic cross in front of the implantation detector, downstream from ARIS. In the same cross, a silicon (PIN) detector recorded the energy loss of each ion as it passed through the detector. Shown in Fig.\ \ref{fig:fig0} is the particle identification (PID) determined from ion energy loss and TOF. In this beam setting, the isotopes $^{22}$N, $^{24}$O, $^{25}$F and $^{26}$F were identified and had significant transmission into FDSi \cite{fdsi}. At the first focal plane of FDSi, the beam was stopped by a YSO (yttrium orthosilicate, Y$_{2}$SiO$_{5}$) based implantation detector \cite{YSO}, which recorded the position and timing information of the implanted ions as well as their subsequent $\beta$-decay electrons. 

$\beta$-delayed $\gamma$ rays were detected by a subset of the Hybrid Array of Gamma Ray Detectors (HAGRiD), which comprised 10 LaBr$_3$ crystals of 2-inch diameter \cite{HAGRiD}. An efficiency curve for HAGRiD was obtained using the known $\gamma$-intensities, $I_{\gamma}$, following the $\beta$ decays of $^{24}$O and $^{25}$F \cite{24F, 24NeExstate, 25Ne}. In this setup, the efficiency of the full-energy peak at 1 MeV was about 1\%. The neutron TOF spectrum was measured using a double wall of 88 Versatile Array of Neutron Detectors at Low Energy (VANDLE) \cite{VANDLE}, with the closest layer 92 cm from the implantation detector. Following the procedure outlined in Ref.\ \cite{133In}, the efficiency curve and neutron response function used for VANDLE was obtained by matching a Geant-4 simulation \cite{geant4} of the decay of $^{49}$K to experimental data \cite{49K} for this well studied isotope. The resultant neutron detection efficiency at 1 MeV was 19\% \cite{ian45cl}. Further information about FDSi and VANDLE are available in Ref.\ \cite{ian45cl}, as the same setup for those specific detectors was used there.

Observed $\beta$ decays were correlated to implantation events within a radius of 0.7 cm and a time interval of 1 s. The radius was chosen to maximize the ratio of signal to background, and this time interval was taken to be about 10 half-lives of $^{25}$F. A total of 1.42 million $\beta$ decays were correlated with $^{25}$F implantation. Neutrons and $\gamma$ rays were also measured within 1 s of implantation into the YSO, with the spectrum taken from before implantation being subtracted from the spectrum taken from after implantation to remove randomly correlated background. 

\begin{figure} [h]
   \includegraphics[width=0.5\textwidth]{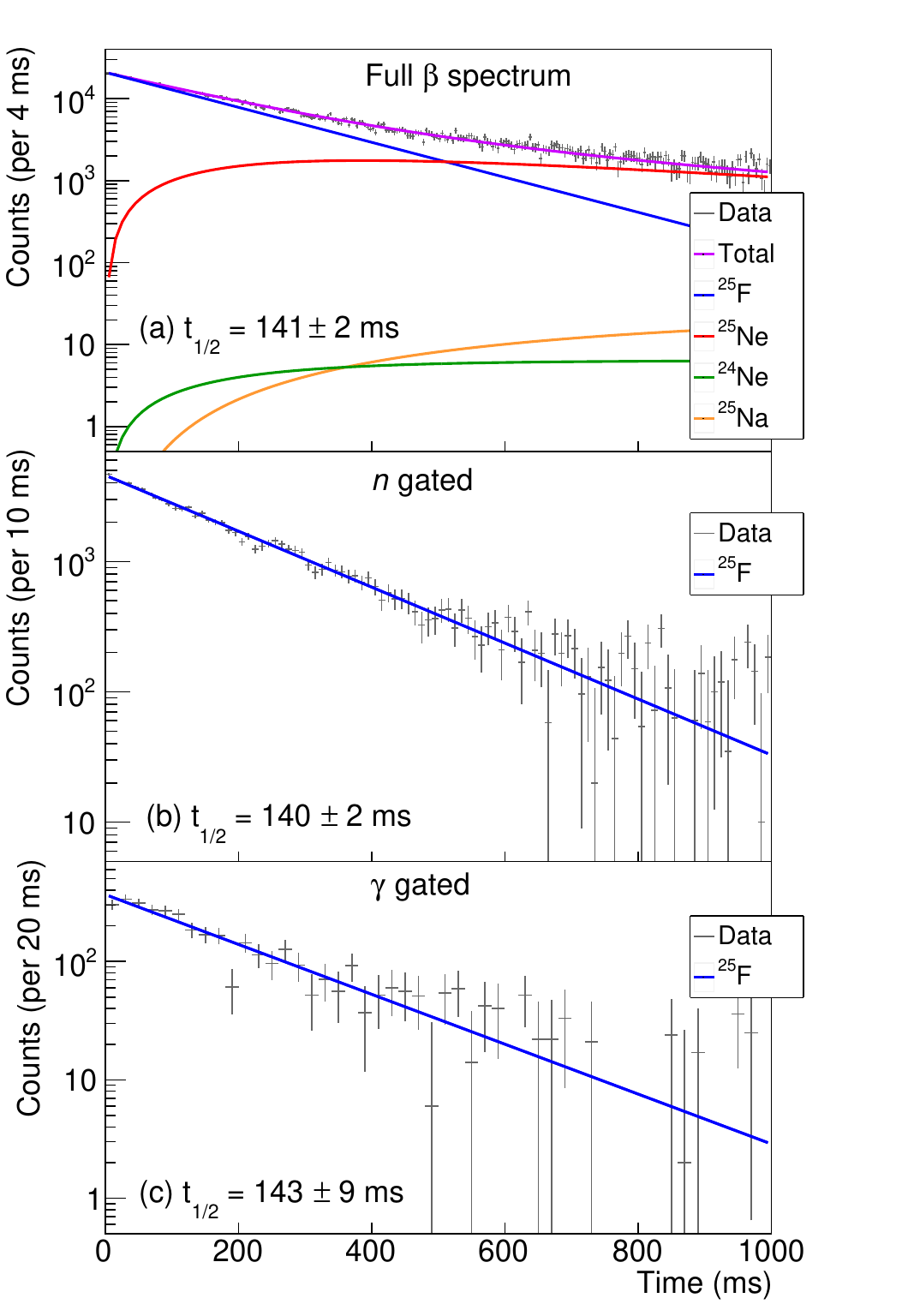}
   \caption{Panel (a) shows the time difference between the implanted $^{25}$F and subsequent $\beta$ decay within the YSO's correlation radius. The purple line is the fitted sum of the activity of $^{25}$F (blue), $^{25}$Ne (red), $^{24}$Ne (green), and $^{25}$Na (yellow) as obtained by the Bateman equations. Panel (b) shows the time difference spectrum gated on neutron detection. The blue line in panel (b) is a single exponential function fit to the data from which the half life is obtained. In panel (c), the time difference between implanted $^{25}$F and detected 1613-, 1622-, or 1702-keV $\gamma$ rays is shown. The blue line in panel (c) is a single exponential fitted to the $\gamma$-gated spectrum, used to find the half life. Corresponding randomly correlated background was subtracted from all of the histograms. The half life obtained from each of these methods is listed in its corresponding panel.}
   \label{fig:fig1}
   \end{figure}

\begin{figure}[ht]
   \includegraphics[width=0.5\textwidth]{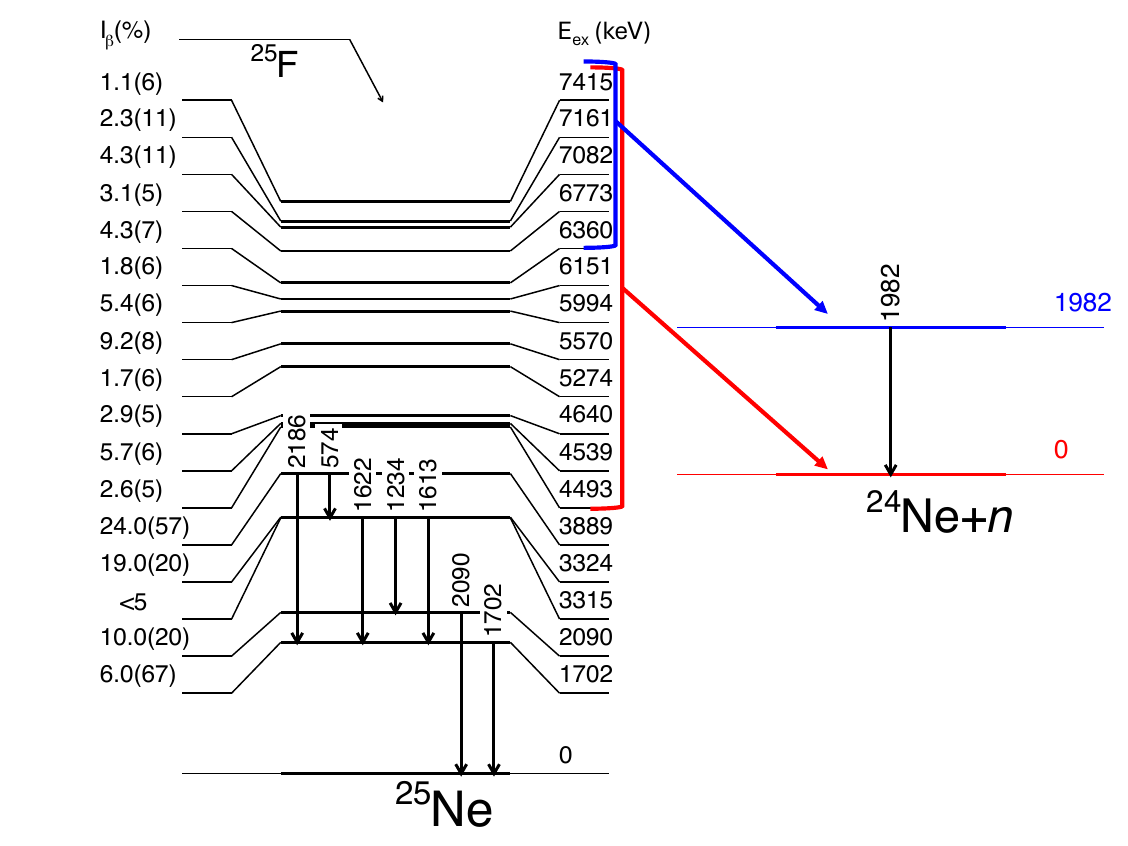}
   \caption{The partial level schemes of $^{25}$Ne and $^{24}$Ne following the $\beta$ decay and $\beta$-delayed neutron emission of $^{25}$F, respectively. $I_{\beta}$ below $S_n$ was taken from Ref.\ \cite{25Ne}, while $I_{\beta}$ above $S_n$ is calculated in the present work. The energy of $^{24}$Ne's first excited state was taken from Ref.\ \cite{24NeExstate}.}
   \label{fig:figlvl}
   \end{figure}

\noindent\textbf{\textit{Results}---}In Fig.\ \ref{fig:fig1}, the half life of $^{25}$F was determined using 3 methods. Figure \ref{fig:fig1}(a) shows the time difference spectrum between implanted $^{25}$F and associated $\beta$ decays, which was fit with a Bateman equation to determine the half life of $^{25}$F. Since the $Q_{\beta}$ of $^{25}$F is only 345 keV above the $S_{2n}$ of
$^{25}$Ne \cite{masstable}, $\beta$-delayed double neutron emission, although possible, is extremely unlikely and thus negligible for the sake of this analysis. Furthermore, $^{25}$F, its daughter and granddaughter from $\beta$ decay ($^{25}$Ne and $^{25}$Na respectively) and its daughter following $\beta$-delayed neutron emission ($^{24}$Ne) were the only nuclei whose activities were accounted for. The reason is as follows: $^{25}$Mg, the daughter of $^{25}$Na, is stable \cite{masstable}, and $^{24}$Na has a long half life (14.955(7) h \cite{24Na}) compared to its parent $^{24}$Ne (3.38(2) min \cite{halflife24Ne}), so its activity was considered negligible as well. The following half-lives were taken from previous studies: $t_{1/2}$ ($^{25}$Ne) = 602(8) ms \cite{halflife25Ne}, $t_{1/2}$ ($^{25}$Na) = 59.1(6) seconds \cite{25Na} and $t_{1/2}$ ($^{24}$Ne) = 3.38(2) minutes \cite{halflife24Ne}. The systematic error on the parent half life was deduced from the uncertainty of the descendants' half-lives from literature and the $P_{n}$ value used in the Bateman equation, which was varied within 44 $\pm$ 2\% (as obtained by the analysis of the neutron spectrum, to be elaborated on later). The resulting half life of $^{25}$F is 141(2) ms, with a statistical and systematic errors of 0.8 and 1.5 ms, respectively. Figure \ref{fig:fig1}(b) shows the decay curve of $\beta$ events in coincidence with the neutron detection. The fit with a single exponential function gave $t_{1/2}$ = 140(2) ms, within 1$\sigma$ of the Bateman equation fit result. In Fig.\ \ref{fig:fig1}(c), the $\beta$-time spectrum was gated in coincidence with a 1613-, 1622-, or 1702-keV $\gamma$ ray, all emitted  from the excited states in $^{25}$Ne following the $^{25}$F decay \cite{25Ne}. The corresponding $\gamma$ transitions in $^{24,25}$Ne are shown in Fig.\ \ref{fig:figlvl}. The $\beta$ spectrum in Fig.\ \ref{fig:fig1}(c) was gated on 3 separate $\gamma$ rays in order to increase the amount of statistics. Fitting the data with a single exponential function, the half life of $^{25}$F was determined to be 143(9) ms, within 1$\sigma$ of both of our other measurements. Our half life is longer than the previously recorded values of $t_{1/2}$ = 50(6) ms \cite{25Fdecay}, 70(10) ms \cite{25FdecayGANIL}, and 90(13) ms \cite{25Ne}. To verify our analysis approach, we applied the Bateman equations to determine the half-lives of $^{22}$N, $^{24}$O and $^{26}$F in the same beam setting, obtaining $t_{1/2}$ = 22.8(2) ms, 138(12) ms, and 9.1(2) ms, respectively, with the error on each value representing the systematic and statistical error added in quadrature. The half life of $^{22}$N has been recorded as $t_{1/2}$ = 21(7) ms \cite{22N}, while the half life for $^{26}$F has been measured as $t_{1/2}$ = 10.2(14) ms \cite{25Fdecay} and 7.8(5) ms \cite{26F} in separate experiments. Along with these, a recent measurement of the half life of $^{24}$O was $t_{1/2}$ = 126(4) ms \cite{PhysRevC.110.034323}. These are all within agreement with our results, supporting our analysis and dataset used to extract the half life of $^{25}$F.

\begin{figure}[h]
   \includegraphics[width=0.5\textwidth]{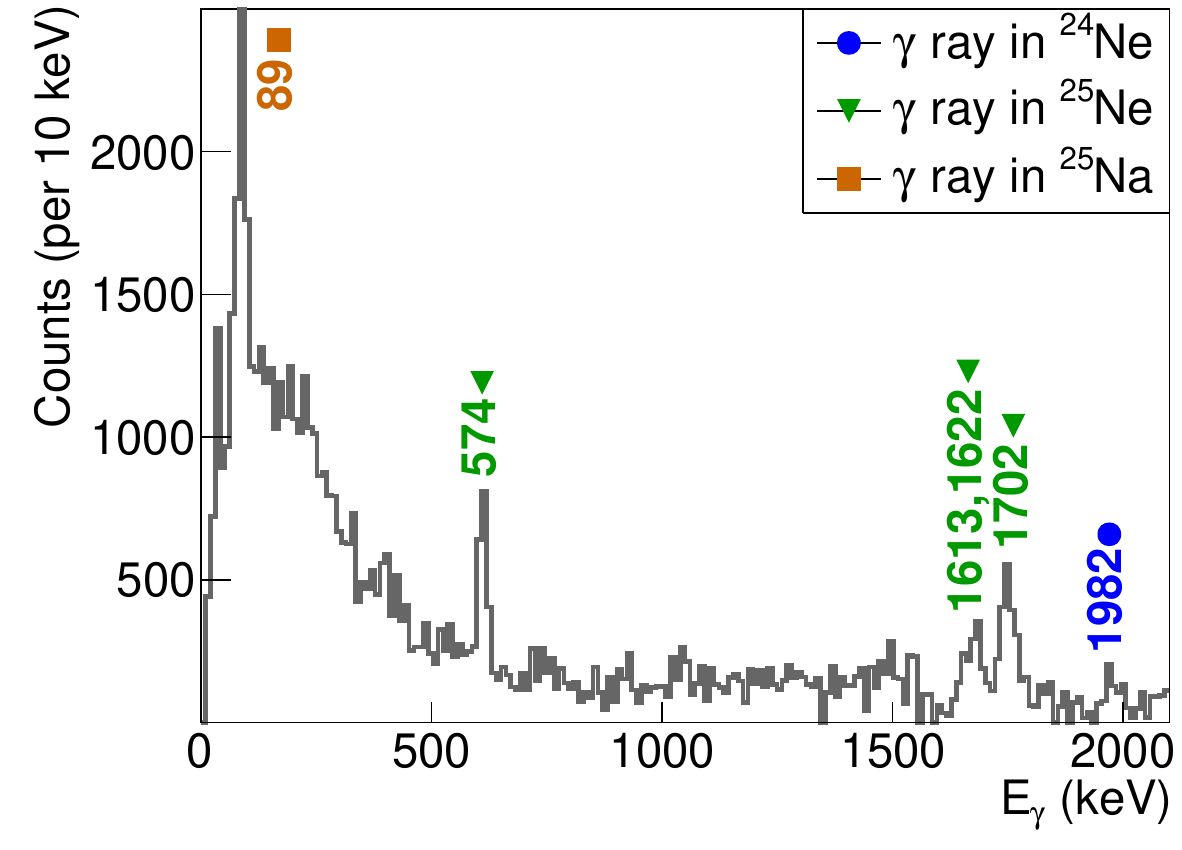}
   \caption{The $\gamma$-ray spectrun using part of the LaBr$_3$ (HAGRiD) array measured within 1 s of $^{25}$F implantation. Background from random correlation was sampled from $\gamma$ rays emitted preceding each implantation, and was subtracted from the histogram.}
   \label{fig:fig_1}
   \end{figure}

Figure \ref{fig:fig_1} displays the $\beta$-delayed $\gamma$ ray spectrum of $^{25}$F taken by HAGRiD. Randomly correlated background was measured before the $^{25}$F implantation and subtracted. Also observed in Refs. \cite{25Ne,25Na, 24NeExstate}, the $\gamma$ rays deexciting the low-lying states in $^{24,25}$Ne and $^{25}$Na are clearly visible, further confirming the ion-$\beta$ correlations are working properly. The 2090-keV peak reported in Ref. \cite{25Ne} is not observed here due to low statistics.

\begin{figure}[ht]
\includegraphics[width=0.5\textwidth]{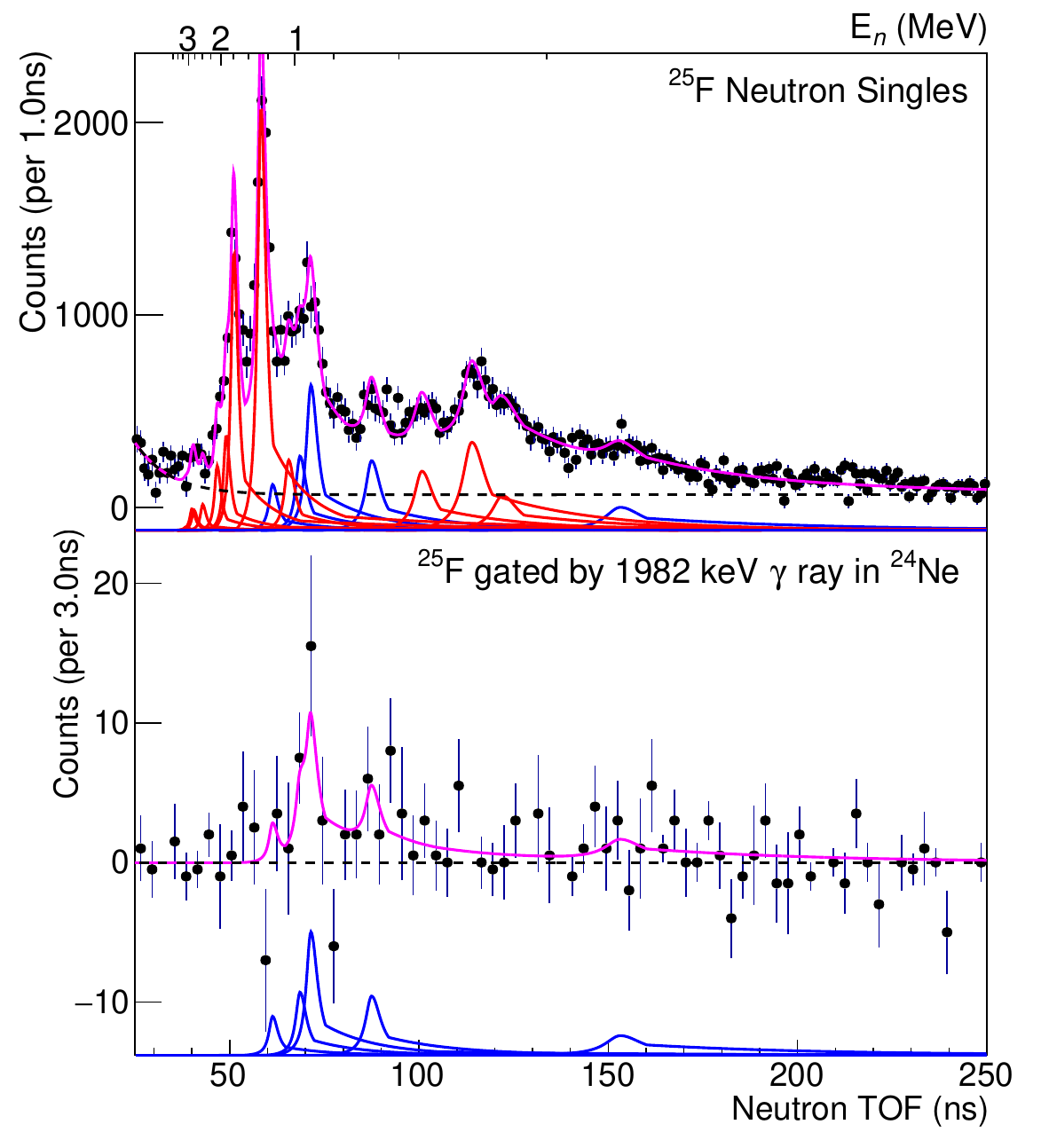}
\caption{Neutron TOF spectrum for $^{25}$F (black), with the full spectrum without any gates applied on the top, and the neutron spectrum taken in coincidence with the detection of a 1982-keV $\gamma$ ray on the bottom. The peaks in the data were fit with a neutron response function, resulting in the red and blue fit lines. The red fit lines represent feeding to the ground state of $^{24}$Ne, and the blue fit lines represent feeding to the first excited state of $^{24}$Ne, which decays to the ground state via.\ emission of 1982-keV $\gamma$ rays \cite{24NeExstate}.}
\label{fig:fig2}
\end{figure}

The TOF spectra for neutrons taken in coincidence with the $\beta$ decay of $^{25}$F are shown in Fig.\ \ref{fig:fig2}. The full spectrum for $\beta$-gated neutrons, neutron singles, is shown in the top panel, while the bottom panel has the spectrum gated on the detection of the 1982-keV $\gamma$ ray emitted during the de-excitation of the first excited state in $^{24}$Ne \cite{24NeExstate}. Decay populating excited states above 1982 keV in $^{24}$Ne is too weak to be observed, and thus was not accounted for in this analysis. A deconvolution based on the method in Ref.\ \cite{PhysRevLett.133.042501} was performed by fitting both the neutron singles and $\gamma$-gated neutron spectra simultaneously with the neutron response functions. In Fig.\ \ref{fig:fig2}, the neutrons feeding the ground and first excited state of $^{24}$Ne are shown as the red and blue peaks, respectively. 

\begin{center}
\begin{figure*}[ht]
   \includegraphics[width=\textwidth]{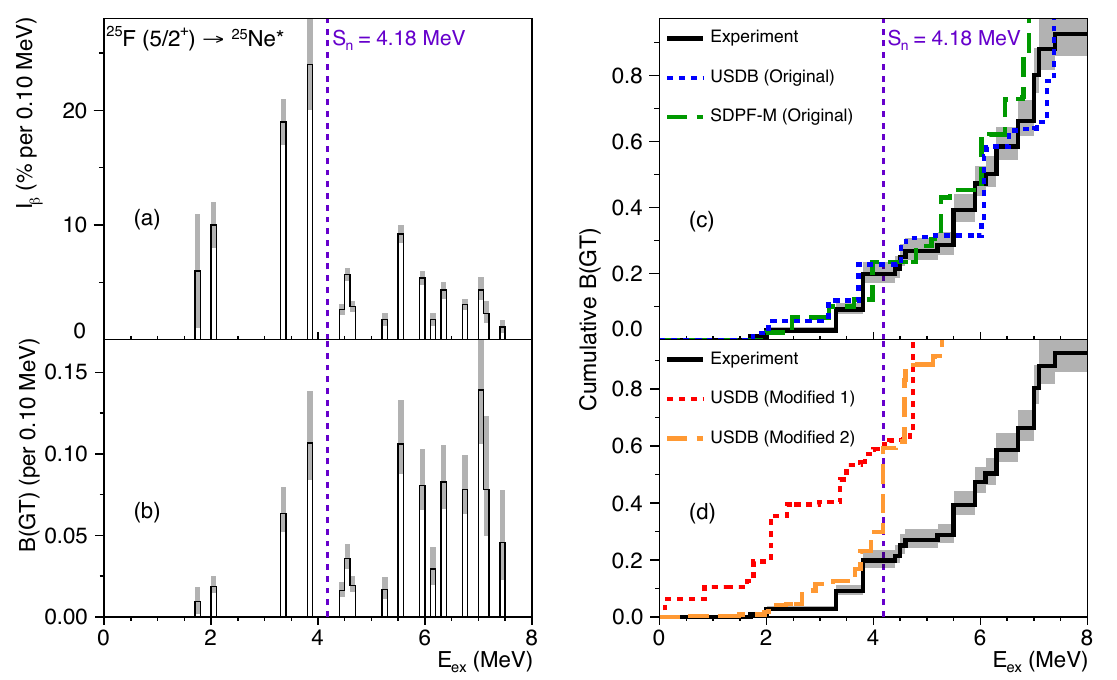}
   \caption{Panel (a) shows $I_{\beta}$ as a function of excitation energy in $^{25}$Ne. Values from below $S_{n}$ were taken from Ref.\ \cite{25Ne}; values above $S_{n}$ were determined from this work. A purple dotted line was placed at $S_{n}$ in $^{25}$Ne, which was obtained from Ref.\ \cite{25Ne}. In panel (b), GT transition strengths, $B$(GT), of $^{25}$F $\beta$ decay are plotted against excitation energy in the daughter $^{25}$Ne. Panels (c) and (d) compare the experimental cumulative $B$(GT) with theoretical predictions. The analyzed experimental data is shown in black, with its error being shown by the light gray shaded region. In panel (c), the shell-model predictions with the USDB \cite{usdbshell} and SDPF-M \cite{sdpfm} interactions are shown in blue and green, respectively. In panel (d), the calculations with two different modifications of the USDB interactions are presented. See text for details.}
   \label{fig:fig3}
   \end{figure*}
\end{center}
   
In Fig\ \ref{fig:fig3}(a), the $\beta$-feeding intensity, $I_{\beta}$, was plotted as a function of excitation energy. For the states below the neutron separation energy of $^{25}$Ne, we took the $I_{\beta}$ from Ref.\ \cite{25Ne} due to its better quality of data measured by HPGe detectors. For the states above $S_n$ in $^{25}$Ne, the values were obtained by the deconvolution discussed above. By normalizing the total number of neutron emissions with the total number of detected $\beta$, the probability of $\beta$-delayed neutron emission, $P_{n}$, was found to be 44(2)\%. Our value of $P_{n}$ is higher than previous measurements: 23.1(45)\% \cite{25Ne} and 15(10)\% \cite{25Fdecay}. In Ref.\ \cite{25Ne}, the $P_n$ value was estimated based on the intensity of the 1982-keV $\gamma$ ray and, therefore, was insensitive to the neutrons feeding the $^{24}$Ne ground state. In Ref.\ \cite{25Fdecay}, the neutron branching ratio was measured with neutron and $\gamma$ detectors independently. While the intensity of the 1982-keV $\gamma$ peak, 11(3)\%, was regarded as the lower limit of the $P_n$ value, the neutron counters gave 15(10)\%, indicating a small fraction of about 4\% feeding the $^{24}$Ne ground state. In our analysis, the fraction of neutron emissions feeding the 1982-keV state in $^{24}$Ne is consistent with that in Ref.\ \cite{25Fdecay}. In contrast, we observed a much stronger feeding to the ground state; see the red peaks in Fig.\ \ref{fig:fig2}. For each state populated in $^{25}$Ne, the partial half life, $t = t_{1/2}/I_{\beta}$, was extracted. The Fermi function $f$ was calculated using $Q_{\beta}=13420(90)$ keV for $^{25}$F \cite{masstable}. Then, the GT transition strengths were given as $B$(GT)$=K/(ft)$, where $K$ = 6144(2) s \cite{bgt}. In Fig.\ \ref{fig:fig3}(b), the experimental $B$(GT) for $^{25}$Ne and their corresponding errors are plotted against excitation energy, $E_{ex}$.

\noindent\textbf{\textit{Discussion}---}In Fig.\ \ref{fig:fig3}(c)-(d), the experimental $B$(GT) distribution was compared with various SM calculations obtained using KSHELL \cite{kshell}. It is noted that the quenching factor of the GT operator is set to 0.75 for all the calculations discussed below. We start from the the USDB effective interaction, which restricts both protons and neutrons to the $sd$ shell \cite{usdbshell}. For almost all $^{25}$Ne excited states below our detection limit of 8 MeV, the USDB interaction predicts their $B$(GT) extremely well. Interestingly, calculations using the USDB interaction also provided a good agreement with observed $B$(GT) of the $^{24}$O decay \cite{PhysRevC.110.034323}. This characterizes the shell structure transition between $^{24}$O and $^{25}$F to be within the prediction of the USDB interaction.

In contrast to this finding, in a proton knockout reaction done on $^{25}$F \cite{25Fvs24O, 25Fpknockout}, the USDB interaction failed to reproduce the spectroscopic factors connecting the ground state of $^{25}$F to the ground and excited states in $^{24}$O. Tang et al.\ suggested a modification of the USDB interaction by reducing the neutron $d_{3/2}$ single-particle energy from 2.1 to -1.5 MeV in $^{25}$F, bringing SM predictions for the spectroscopic factors into agreement with their experimental data \cite{25Fvs24O}. According to the effective single-particle energies, the $N=16$ shell gap in $^{24}$O is 5.4 MeV, whereas that in $^{25}$F is 4.8 and 1.2 MeV before and after the modification, respectively. When this modified USDB interaction is employed to calculate the low-lying states in $^{25}$Ne, their energies severely contradict the experimental data. In particular, the first excited state, which is predicted to be 3/2$^+$, appears at 113 keV, whereas the experimental first excited state is at 1.7 MeV and has a spin-parity assignemnt of 5/2$^+$ \cite{Catford:2010sz}. The resulting $B$(GT) distribution is thus shifted significantly to lower excitation energy, shown as ``Modified 1'' (the red line) in Fig.\ \ref{fig:fig3}(d). This starkly disagrees with our experimental $B$(GT), which is more consistent with the nominal calculations.

In addition, we considered a second modification of the USDB interaction. Instead of reducing the single-particle energy of the $0d_{3/2}$ orbital directly, we change the monopole interactions between the proton $0d_{5/2}$ and neutron $0d_{3/2}$ orbitals. To reproduce the spectroscopic factors reported in Ref.\ \cite{25Fvs24O}, we increased the amplitudes of these monopole interactions by 4 MeV. Then, we used the interaction to calculate the $B$(GT) from $^{25}$F to $^{25}$Ne, which is shown as ``Modified 2'' (the orange line) in Fig.\ \ref{fig:fig3}(d). In contrast with ``Modified 1'', this second modification agrees with the experimental data below $S_n$. However, it gives 5/2$^+$ as the ground-state spin and parity of $^{25}$Ne, whereas the experimental assignment is 1/2$^+$ \cite{FERNANDEZDOMINGUEZ2007389}. Furthermore, the modified interaction fails to reproduce the $B$(GT) distribution above $S_n$. As such, this interaction cannot give consistent interpretations for the $^{25}$F(p, 2p)$^{24}$O reaction \cite{25Fpknockout} and the $^{25}$F $\beta$ decay. From the $\beta$ decay perspective, the evolution of the neutron $0d_{3/2}$ orbital in $^{25}$F and $^{25}$Ne is best characterized by the original USDB interaction, suggesting the $N = 16$ shell gap is mostly preserved for $^{25}$F. A similar conclusion has been drawn in a recent work by Neupane et al.\ for the $^{24}$O decay \cite{PhysRevC.110.034323}.

The USDB interaction only takes into account the proton and neutron $sd$ shell \cite{usdbshell}. Therefore, it will not be able to describe nucleon excitation to higher energy shells, such as the $pf$ shell. In order to investigate whether such cross-shell excitations play an important role in the $^{25}$F decay, we performed the SM calculations using the SDPF-M interaction, which includes the full $sd$ shell as well as the $0f_{7/2}$ and $1p_{3/2}$ orbitals \cite{sdpfm}. In Fig.\ \ref{fig:fig3}(c), the green dashed line represents the cumulative $B$(GT) as predicted by the SDPF-M interaction. The calculated results match the experimental data nearly as well as those using the USDB interaction. The difference between these two interactions is very small, implying that with the current level of experimental error, both of these interactions well describe observables. Along with the changes in the $N = 16$ shell gap, a larger $N = 20$ shell gap is predicted for $^{25}$F than for $^{24}$O \cite{unstable_magic_number, SM_evolution_exotic_nuclei}. In the SDPF-M calculation, the average occupancy of the neutron $pf$ shell was $\sim$0.04 at the 1/2$^+$ ground state of $^{25}$Ne, and increases to only $\sim$0.1 in the 3/2$^+$, 5/2$^+$, and 7/2$^+$ states at 6 MeV in $^{25}$Ne. Given this model's agreement to our data, it is concluded that the cross-shell excitation is not significant in the $\beta$ decay of $^{25}$F, likely a consequence of a large $N = 20$ shell gap in $^{25}$F. This could be the reason SM calculations employing the USDB interaction, in which neutron orbitals above $N = 20$ are not included, are produce even better agreement for the $B$(GT) of $^{25}$F than for $^{24}$O as shown in Ref.\ \cite{PhysRevC.110.034323}.

\noindent\textbf{\textit{Conclusion}---}The $\beta$-delayed neutron emission of $^{25}$F was studied in detail; for the first time the transitions to the neutron unbound states in $^{25}$Ne were measured. The half life and $P_{n}$ values of $^{25}$F were remeasured to be $t_{1/2}$ = 141(2) ms and $P_{n}$ = 44(2)\%, both of which are higher than past measurements. The $B$(GT) values were compared to several SM calculations with different effective interactions. Predicted $B$(GT) from the USDB and SDPF-M interactions were in very close agreement to experiment for most states in $^{25}$Ne whereas modifications that bring predicted spectroscopic factors into agreement with experiment fail to reproduce the $\beta$-decay results. Comparing the USDB and SDPF-M calculations revealed cross-shell excitation is not significant in the $\beta$ decay of $^{25}$F. The close agreement of SM calculations employing the USDB interaction to experimental results implies that in contrast to past works (Refs. \cite{25Fvs24O, 25Fpknockout}), the $N = 16$ shell gap is largely preserved in $^{25}$F, only reduced by 0.6 MeV upon the addition of a proton from $^{24}$O. This emphasizes the importance of complentary studies on all obtainable observables to understand nuclear structure. Thus, the lack of spectroscopic overlap between the ground state of $^{25}$F and the ground state in $^{24}$O remains unexplained, and a further study on $^{25}$F is needed to find a solution to explain the discrepancy between different experimental approaches.

\section*{Acknowledgments}
We would like to thank the entire operations team at FRIB for reliable beam delivery during the experiment. We would also like to thank Dr.\ Ben Kay, Dr.\ Kate Jones, Dr.\ Fr\'ed\'eric Nowacki, and Dr.\ Tsz Leung Tang for insightful discussions. This work is supported in part by the U.S. Department of Energy, Office of Science, Office of Nuclear Physics under Contracts No. DE-AC52-07NA27344 (LLNL), No. DE-AC05-00OR22725 (ORNL), No. DE-FG02-96ER40983 (UTK), No. DE-SC0020451 (MSU), Award No. DE-SC0020451(FRIB), and used resources of the Facility for Rare Isotope Beams (FRIB) Operations, which is a DOE Office of Science User Facility under Award Number DE-SC0023633. This work was also sponsored by the Stewardship Science Academic Alliances program through DOE Awards No. DE-NA0003899 (UTK) and No. DOE-DE-NA0003906 (MSU), NSF Major Research Instrumentation Program Award No. 1919735 (UTK), the US National Science Foundation under grants No. PHY-20-12040 (B.M.S) and PHY-1848177 (CAREER), the Department of Energy National Nuclear Security Administration through the Nuclear Science and Security Consortium under Award Number DE-NA0003996, and the U.S. Department of Energy, National Nuclear Security Administration under award No. DE-NA0003180 (MSU).

% \bibliographystyle{model1-num-names}
%\bibliographystyle{elsarticle-names}
% \bibliography{ref}

\end{document}